\documentstyle[preprint,aps]{revtex}
\input psfig
\begin{document}
\draft
\title{Exact eigenstates and transmission for two interacting electrons\\
 on quantum dots}
\author{Amnon Aharony and O. Entin-Wohlman}
\address{\it School of Physics and Astronomy, Raymond and Beverly Sackler 
Faculty of Exact Sciences, \\
Tel Aviv University, Tel Aviv 69978, Israel \\ }
\author{Y. Imry and Y. Levinson}
\address {\it Condensed Matter Physics, The Weizmann Institute of Science,
Rehovot 76100, Israel}
\date{\today}
\maketitle
\begin{abstract}
The eigenstates and the scattering transmission for two interacting electrons
are found exactly for $I$ quantum dots, including the hybridization
with the states on the leads.
The results imply limitations on the validity of the
Coulomb blockade picture.
The ground states for $I=1,\ 2$
on a one--dimensional chain (modeling single and double quantum dots)
exhibit quantum delocalization
and magnetic transitions.
The effective transmission $T$ of two interacting electrons through 
one impurity ($I=1$)
is enhanced by a renormalization of the repulsive interaction, when
one of the electrons is captured in a strongly localized state.
\end{abstract}

\section{Introduction}
\label{intro}

There has been much recent interest in the effects of interactions
on the localization of electrons in disordered systems \cite{belitz}.
Following many recent papers, we address this issue for the simplest
case of {\it two electrons}.
Recent numerical work 
\cite{shepelyansky,vonoppen,ortuno} found that interactions help 
to delocalize the electrons, yielding a metal--insulator transition.
Instead of discussing a fully random system, we consider a
{\it dilute} system with only $I$ (out of $N$) impurities.
We then discuss the `static'
eigenstates and scattering situations.
In Sec. 2 we show that the exact eigenenergies of the two electrons, 
in a tight binding model with on--site interactions,
are found
from the eigenvalues of a (small) $I \times I$ matrix, which involves the
solutions of the 1e
Hamiltonian. This yields a new calculational scheme, which should be
very effective for
numerical studies.

Using this scheme, we obtain  consequences for
{\it quantum dots} \cite{review}, and our discussion emphasizes the 
importance of the hybridization of the wave functions with the `band' 
states of the `leads'.
Quantum dots have potential applications as
artificial atomic or molecular
devices, and are very instrumental in studying
strong correlations and their implications.
Much of the discussion in the literature used the Coulomb blockade approach 
\cite{coulomb}, which asigns a constant repulsive energy for each pair
of electrons on the dot.
In contrast, we find that the interacting eigenenergies
are bounded between consecutive non--interacting 
two--electron (2e) energies, so that
the energy cost due to the interactions is usually much smaller than the
average on--site repulsion $\langle U \rangle$.

In Sec. 3 we apply our scheme to specific models of quantum dots.
A single quantum dot coupled to electrodes is
modeled by one
impurity ($I=1$) on a one dimensional (1D), $N$--site chain
\cite{ng}. This represents a special case of
the Anderson model \cite{anderson},
with a specific momentum--dependent hybridization between the impurity and 
the conductance band. 
We find that the behavior of the two electrons
on the ``dot" has a rich phase diagram, as function of
the ``dot" site energy $\epsilon_0$
and the hybridization.
These parameters can be tuned experimentally, by varying the
voltage on a gate coupled capacitively to the dot and the barriers between
the dot and the leads \cite{dotexp}.
We also find a delocalization transition at large $U$, for sufficiently 
negative $\epsilon_0$, giving an analytic demonstration of the metal--insulator
transition mentioned above.
More complex examples include a cluster of $I$ impurities coupled to
1D leads,
representing one large dot \cite{berkowitz}, 
two separate impurities, representing
double quantum dots (or artificial ``molecules") \cite{double}, etc.

Section 4 generalized the above discussion to scattering
situations.
For a simple model, with $I=1$, we find a large enhancement of the 
transmission by the interactions, reflecting again delocalization by
interaction. We find peaks in the transmission (or in the conductance)
as function of the gate voltage on the dot. These peaks have nothing to 
do with those discussed in the Coulomb blockade literature.

\section{Scheme for $I$ impurities}

We start with the 1e
tight binding Hamiltonian
\begin{equation}
{\cal H}_0= \sum_{\langle n,m \rangle}(t_{nm}|n \rangle\langle m|
+h.c.)+ \sum_{i=1}^I \epsilon_i |i \rangle \langle i|,
\label{H0}
\end{equation}
where $|i \rangle$ is a (spin--independent) state localized
on site $i$, the first sum runs over
all the site pairs in the system (including the impurities) and the second
sum runs over the $I$ impurities.
We first find the 1e eigenstates 
$|a \rangle \equiv \sum_n \phi_a(n)|n \rangle$ 
and eigenenergies $\epsilon_a$ of ${\cal H}_0$. 
For the interactions, we consider only {\it on--site repulsive interactions},
${\cal H}_{\rm int}= \Sigma_{i=1}^I 
U_i\hat n_{i,\uparrow} \hat n_{i,\downarrow}$.
This interaction has no effect on the {\it triplet} states, which vanish
when both electrons are on the same impurity.
Therefore, we discuss only {\it singlet} eigenstates, $\Psi(i,j)=\Psi(j,i)$,
with electrons at sites $i$ and $j$.

Rewriting the eigenvalue equation $({\cal H}-E)\Psi(i,j)=0$ 
in the form $\Psi=(E-{\cal H}_0)^{-1}{\cal H}_{\rm int}\Psi$,
with ${\cal H}={\cal H}_0+{\cal H}{\rm int}$, we
can now easily obtain the equations $A_i= \Sigma_{j=1}^I S_{ij}A_j$,
where
$A_i \equiv \sqrt{U_i}\Psi(i,i)$, and $S_{ij}/\sqrt{U_iU_j}=G^0_E(ii,jj)
\equiv \langle ii|(E-{\cal H}_0)^{-1}| jj\rangle$
is the non--interacting
2e Green's function in which the two electrons are on the same site,
cf. \cite{vonoppen}.
The new eigenvalues are thus found by requiring that
the $I \times I$ determinant $D(E) \equiv ||S_{ij}- \delta_{ij}||$ vanishes.
Finding the
$I$ eigenvalues ${\cal S}_i(E)$ of the matrix $S$,
the new ($U_i$--dependent)
exact eigenenergies $\{E\}$ solve the equations ${\cal S}_i(E)=1$.
The $A_i$'s are then given by the 
eigenvectors related to ${\cal S}_i(E)$,
and $\Psi(m,n)=\Sigma_iG_E^0(mn,ii)\sqrt{U_i}A_i$.

Using the non--interacting singlet basis,
$\Psi^{ab}_0(i,j)=
[\phi_a(i)\phi_b(j)+\phi_a(j)\phi_b(i)]/2^{(1+\delta_{ab})/2}$,
with eigenenergies $\epsilon_{ab}=\epsilon_a+\epsilon_b$ 
and $\epsilon_a \le \epsilon_b$,
one has
\begin{equation}
G^0_E(ij,k\ell)=\sum_{ab}\phi_a(i)\phi_b(j)\phi_a^\ast(k)\phi_b^\ast(\ell)/
(E-\epsilon_{ab}).
\label{G00}
\end{equation}
Since each $S_{ij}$ has poles at every $\epsilon_{ab}$, 
each ${\cal S}_i$ will generally also have such poles.
Ignoring special symmetric cases, when the residues of some of these
poles vanish,
one has 
$S_{ij} \propto
1/(E-\epsilon_{ab})$ when $E \approx \epsilon_{ab}$. 
In this approximation, $D(E) \approx
(-1)^I(1-{\rm Tr} S) \approx (-1)^{I} [1-\langle U \rangle_{ab}
/(E-\epsilon_{ab})]$,
with $\langle U \rangle_{ab} \equiv \sum_i U_i |\Psi_0^{ab}(i,i)|^2$. 
As $E$ crosses $\epsilon_{ab}$, $(-1)^{I+1}D(E)$ jumps from $-\infty$
to $\infty$. As $E$ increases between two consecutive non--interacting 
eigenvalues, $(-1)^{I+1}D(E)$
varies smoothly from $\infty$ to $-\infty$.
Thus, $D(E)=0$ must have at least one (and up to $I$) solution(s)
between every pair
of such consecutive energies, and the new energies maintain the sequence
of the non--interacting ones. For example,
the cost of adding two electrons into the 
ground state, $\Delta_{gg}=E_{gg}-\epsilon_{gg}$, is thus bounded by
the 1e distance to the first excited level, 
$\epsilon_u-\epsilon_g$, which may be much smaller than the Coulomb blockade
value $\langle U \rangle$!
It is interesting to note that small (and not evenly spaced)
values of $\Delta$ were observed in
a series of 2D quantum dots \cite{ashoori}, apparently contradicting the 
simple Coulomb blockade picture.
It is tempting to relate these observations
to our result.

\section{One and two impurities: ground state and phase transitions}

\begin{figure}
\vspace{-1.5cm}
 \leftline{\psfig{figure=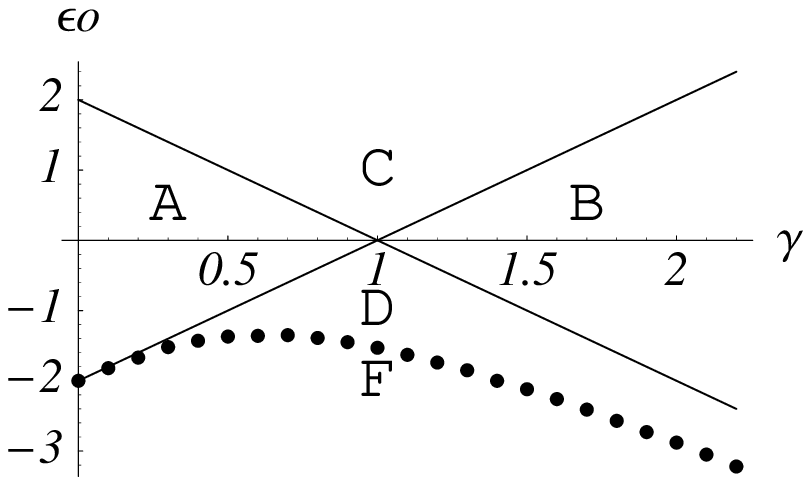,width=13cm}}
\vspace{-13.5cm}
\caption{$\epsilon_0-\gamma$ phase diagram for the single impurity case.}
\label{fig1}
\end{figure}

The simplest example concerns one impurity (or ``dot") on a 1D wire, closed with
periodic boundary conditions. We place the impurity at site $i=0$, with
energy $\epsilon_0$ and with real matrix elements $t_{0,1}=t_{0,N-1}
\equiv t_0$.
The other nearest neighbor matrix elements are set at $t_{n,n+1}=t$,
for $n=1, 2, ...,N-2$.
The 1e spectrum contains $N/2$ odd eigenstates, $\phi_k(n) \propto \sin(kn)$,
which are not affected by the impurity, and $N/2$ even ones, whose energies
are shifted.
These even  solutions exhibit a rich phase diagram
in terms of $\epsilon_0$ (in units of $t$) and $\gamma \equiv (t_0/t)^2$,
as shown in Fig. \ref{fig1} \cite{aharony}:
In region A all the states are delocalized, with almost unshifted
energies, in the `conduction band'.
In region C
(or D+F) there exists one bound state above (or below) the conduction
band.
Finally, both bound states exist in region B.
An important detail concerns the normalization: except for the trivial
case when $\gamma=1$ and $\epsilon_0=0$, the amplitude of the `band'
state $|\phi_k(0)|^2$ is found to be proportional to
$\sin^2k$, vanishing at the band edges.
This will turn out to
be crucial below.


For $I=1$,
adding the interaction $U_1 \equiv U$, $D(E)=S_{11}(E)-1$.
Indeed, $S_{11}(E)$ jumps from $-\infty$ to $\infty$ as $E$ crosses
each non--interacting energy $\epsilon_{ab}$,
and we find one new eigenvalue between every pair of non--interacting
energies, as described above. In regions B, D and F, 
the non--interacting ground state has two bound electrons, with
energy $\epsilon_{gg}=-2|\epsilon_g|$, and the new ground state energy is always
bound between this energy and that of one bound and one `free' electron,
$\epsilon_{gu}=-|\epsilon_g|-2t$.
For finite $N$, $S_{11}$ would diverge towards $-\infty$
as $\epsilon_{gu}$ is approached, implying a persistent doubly
bound solution
below the band for all $U$.
However, when $N \rightarrow \infty$ we replace the sums over the band states
by integrals.
Since $|\phi_k(0)|^2$ vanishes at $\epsilon_{gu}$,    
$S_{11}(\epsilon_{gu}) \equiv Us_c$ remains {\it finite}.
In region D one has $s_c<0$, so that $D(E)=0$ has a discrete doubly
bound state for all $U$. However, in region F one has
$s_c>0$,
implying a disappearance
of this bound state for
$U>U_c=1/s_c$; the ground state energy is now at the bottom of the band,
implying an `ionization' of one electron and an `insulator to metal'
transition from D to F. 
This transition occurs at a lower value of $U$ for larger
negative values of $\epsilon_0$,
when the localization length of the bound
electrons is smaller, and they feel the interaction more strongly.

It should be noted that although the bound state is a singlet, 
with total spin zero, the new ground
state in region F has one bound electron and one ``free" electron. Such a
state does not feel the e--e repulsion, and is thus practically 
degenerate with the slightly lower triplet state (for large $N$, the difference
is of order $1/N$). 
Unlike the ``insulator" singlet (or ``antiferromagnetic")
ground state, which has no net magnetic moment,
this ``metallic" state in region F is {\it paramagnetic}. 
This difference should
be measurable in an external magnetic field.
Our model also exhibits interesting finite size effects for $N<\infty$,
and interesting antibound 2e excitations above the band \cite{aharony}.

The analysis is somewhat more complicated for $I=2$, where
the $2 \times 2$ matrix $S$ has symmetric and antisymmetric eigenvalues,
corresponding to products of even--even (and odd--odd) and of even--odd
1e states, respectively. 
The resulting phase diagram now also depends on the inter--impurity
distance $R$, and the region equivalent to F in Fig. \ref{fig1} shrinks
to a small `bubble' \cite{aharony}.

\section{One impurity: transmission}

So far we emphasized only the ground state, for a problem with periodic
boundary conditions. In real `quantum dot' experiments one is more
interested in {\it scattering} situations, when electrons are sent
from one side and one calculates the {\it current} or the {\it transmission}
through the dot.
For the simplest exact calculation of the transmission for two
interacting electrons,
we replace the above
tight binding Hamiltonian by its continuum limit,
\begin{eqnarray}
{\cal H}(x_{1},\sigma_{1},x_{2},\sigma_{2})&=&{\cal H}_{0}(x_{1},x_{2})
+U\delta (x_{1})\delta (x_{2})\delta_{\sigma_{1},-\sigma_{2}},\nonumber\\
{\cal H}_{0}(x_{1},x_{2})&=&{\cal H}_{\rm sp}(x_{1})+{\cal H}_{\rm sp}(x_{2}),
\label{H}
\end{eqnarray}
where $x_{i}$  and $\sigma_{i}$ are the coordinate and the spin component of the
$i$th electron, and
${\cal H}_{\rm sp}$ is the single--particle Hamiltonian, 
independent of the spin components.

Again, we consider only the singlet spatially {\it symmetric} wave
functions, $\Psi(x_{1},x_{2})=\Psi(x_2,x_1)$.
At total energy $E$, we split $\Psi$ into
$\Psi(x_{1},,x_{2})
=\Psi_{0}(x_{1},x_{2})+
\Psi_{\rm S}(x_{1},x_{2})$,
where $\Psi_{0}$ is the solution of ${\cal H}_{0}(x_{1},x_{2})$, 
with the same energy $E^+\equiv (E+i\eta )$ (with $\eta \rightarrow 0$),
$({\cal H} _{0}-E^+)\Psi_{0}=0$.
For the on--site interaction (\ref{H}) it then follows that 
$\Psi_{\rm S}(x_{1},x_{2})=
UG_{E}(x_{1},x_{2};
0,0)
\Psi_{0}(0,0)$,
where $G_{E}$ is the two--particle Green's function of the {\it interacting
Hamiltonian}.

For the model Hamiltonian given by (\ref{H}) one can express 
the Green's function
of the interacting system, $G_{E}$, in terms of 
$G_{E}^{0}$, discussed in the previous section.
The definitions of these two Green's functions
yield the Bethe--Salpeter equation
\begin{equation}
G_{E}(x_{1},x_{2};
x_{1}',x_{2}')=G_{E}^{0}(x_{1},x_{2};x_{1}',x_{2}')
+UG_{E}^{0}(0,0;x_{1}',x_{2}')
G_{E}(x_{1},x_{2};
0,0),
\end{equation}
and hence
$G_{E}(x_{1},x_{2};0,
0)=
G_{E}^{0}(x_{1},x_{2},0,0)/[1-UG_{E}^{0}(0,0,0,0)]$.
Thus,
$\Psi
(x_{1},x_{2})=\Psi_{0}(x_{1},x_{2})
+F_E G_{E}^{0}(x_{1},x_{2},0,0)\Psi_{0}(0,0)$,
where 
$F_{E}=U/[1-UG_{E}^{0}(0,0,0,0)]$.
Thus, $\Psi_S$ is determined
solely by the non--interacting Hamiltonian ${\cal H}_{0}$. 

We next calculate 
the quantum average of the current density operator at $x=x_0$,
in the exact singlet state $\Psi$,
\begin{equation}
j(x_{0})=\frac{2e\hbar}{m}
\Im \int dx_{1}dx_{2}\delta (x_{1}-x_{0})
\Psi^{\ast}(x_{1},x_{2})\frac{d}{dx_{1}}\Psi (x_{1},x_{2}).
\end{equation}
The explicit calculation of $j$ now requires only integrals involving
the non--interacting functions $\Psi_0(x_1,x_2)$ and non--interacting
states $\phi_p(x)$.
In what follows, we shall assume that $\Psi_0$ is given by
the singlet combination $\Psi_0^{pq}$, as defined in the previous section,
and that the total energy is given by 
$E=\epsilon_p+\epsilon_q$.

To proceed, 
we choose a simple $\delta$--function attractive
potential,
\begin{eqnarray}
{\cal H}_{\rm sp}(x)=-\frac{\hbar^{2}}{2m}\frac{d^{2}}{dx^{2}} 
-V\delta (x),\label{Hsp}
\end{eqnarray}
which has one bound state, $\phi_b=\sqrt{\kappa}e^{-\kappa |x|}$
with the inverse localization length $\kappa=mV/\hbar^2$, 
with eigenenergy $-\epsilon_b=-\hbar^2\kappa^2/(2m)$,
and ``band" scattering wavefunctions
$\phi_p=(e^{i p x}+r_p e^{i p |x|})/\sqrt{L}$, with the
reflection and transmission amplitudes
$r_p=i \kappa/(p-i \kappa),\ t_p=p/(p-i \kappa)$,
and with eigenenergy
$\epsilon_p=\hbar^2p^2/(2m)$.

There are two physical
situations which are of interest for
the non--interacting wave function $\Psi_0^{pq}$.
The first 
corresponds to two {\it propagating} electrons, impinging from the left,
when both $p=p_1$ and $q=p_2$ represent ``band" states.
In this case, the current density of
a `macroscopic' system (that is, for $L\rightarrow \infty $) is {\it
unaffected} by the
interaction $U$, and remains as in the non--interacting system,
$j=e\hbar (p_{1}|t_{p_{1}}|^{2}+p_{2}|t_{p_{2}}|^{2})/(mL)+O(U/L^2)$.

The second, more interesting, choice for $\Psi_0^{pq}$
arises when one electron is propagating, with $p$ representing its wave vector,
and the
other is captured in the bound state, i. e. $q=b$. 
Now the total energy is
$E(p,b)=\epsilon_{p}-\epsilon_{b} \ge -\epsilon_b$.
In this case, the terms coming from the interaction are of the same order
as the non--interacting ones.
A long but straightforward calculation now yields 
$j=e\hbar p T(p)/(mL)$,
where the effective transmission is
$T(p)=|t_{p}|^2-2m\Re(F_E r_p t_p)/\hbar^2$.
Remembering that $F_{E}=U/[1-UG_{E}^{0}(0,0;0,0)]$,
the new term in $T$ will have a `resonance' when 
$\Re F_E^{-1}=U^{-1}-\Re G_E^0(0,0;0,0)=0$. This equation is similar to 
the equation $S_{11}=1$ encountered in the previous section.

\begin{figure}
\centerline{\psfig{figure=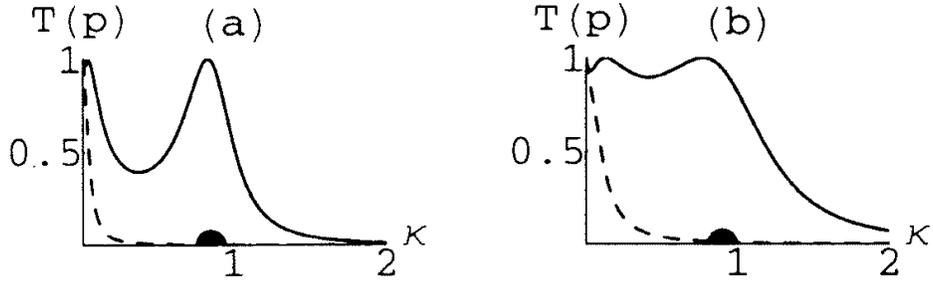,width=14cm}}
\caption{The
transmission $T$ as function of $\kappa $ (in units of the cutoff $\omega$),
for $U=5\hbar^2/(2m)$.
The solid line is
$T(p)$, the dashed line shows $|t_{p}|^{2}$. (a) $p=0.04\omega$,
(b) $p=0.1\omega$.}
\label{fig11}
\end{figure}

For an explicit calculation of $G_{E}^{0}(0,0;0,0)$, 
we introduce an upper cutoff 
$W=\hbar^{2}\omega^{2}/2m$
on the ``band" states. We then found that
$\Re G_{E}^{0}(0,0;0,0)<0$ for small $\kappa$, and $\Re G_{E}^{0}(0,0;0,0)>0$
for large $\kappa$.
In the former region there is no `resonance'. In the latter region there
can be a resonance for sufficiently large $U$.
This yields a very
interesting dependence of the effective transmission $T$ on
the single--electron parameters and on $U$. Specifically, Fig. \ref{fig11}
shows the dependence of $T(p)$ on $\kappa$, for a finite value
of $U$ and for two values of $p$.
Clearly, $T(p)$
increases significantly as the result of the interaction,
reflecting {\it delocalization due to the interaction}.
At $\kappa=0$ one has $T=1$, independent of $U$. At small $\kappa$,
$T$ always decreases, due to scattering by the single electron potential,
following the non--interacting case.
However, $U$ effectively screens this attractive potential, and
at small $p$ this results in a double peak structure of $T$, as shown
in the figure.
Qualitatively, these peaks arise when the screening cancels the attractive
potential.
At larger $p$ the minimum between the peaks grows, the two peaks join
into a relatively broad plateau which decays at large $p$.

%



This project is supported by the Israel Science Foundation
and by a joint grant from the Israeli Ministry of Science
and the French Ministry of Research and Technology.
O. E-W also thanks the Albert Einstein Minerva Center for 
Theoretical Physics for partial support.


\end{document}